\documentstyle[aps,prl,preprint]{revtex} 
\tightenlines 
\begin{document} 
\bibliographystyle{unsrt} 
\draft 
\title{Static magnetization induced by time-periodic fields 
with zero mean} 
\author{S. Flach and A. A. Ovchinnikov} 
\address{Max-Planck-Institute for the Physics of Complex Systems, 
N\"othnitzer Str. 38,\\
 D-01187 Dresden, Germany  
} 
\date{\today} 

\maketitle 

\begin{abstract}
We consider a single spin in a constant magnetic field or
an anisotropy field. We show that additional 
external
time-periodic fields with zero mean may generate nonzero time-averaged
spin components which are vanishing for the time-averaged Hamiltonian. 
The reason is a lowering of the dynamical symmetry of the
system. A harmonic signal
with proper orientation is enough to display the
effect. We analyze the problem both with and without dissipation,
both for quantum spins ($s=1/2,1$) and classical spins.  
The results are of importance for controlling
the system's state using high or low frequency fields and for 
using new resonance techniques which probe internal system parameters,
to name a few.
\end{abstract} 
 
\pacs{75.10.Jm,76.20.+q,76.60.-k}

Usually nonzero averages of observables, which would be expected
to be zero by symmetry considerations, are generated either by
constant external fields, or by internal interactions
which may lead to phase transitions. However as
we will show below such a situation is also possible if we use
time-periodic fields with zero mean. 
The general idea behind the following results
is purely symmetry related, and thus it seems to be worthwhile
to understand the mechanisms which may lead to nonzero averages
if such fields are applied. 
This work is motivated by a recent paper \cite{sfoyyz00} where
similar ideas have been used to explain the phenomenon of directed
currents in driven systems. 
The essence of the present paper is that we can lower the symmetry
of a given dynamical system by applying time-periodic fields with
zero mean, i.e. that the time-averaged Hamiltonian
displays symmetries which would imply zero averages
for corresponding observables. It will be the symmetry breaking in the
temporal evolution which induces nonzero averages.

Let us start our considerations with a model describing an $s=1/2$ spin
in a constant field $h_z=2$ directed along the $z$-direction and
a time-periodic field $2h_x(t)$ with period $T$ and zero mean directed along the
$x$-direction. The Hamiltonian is given by $H=h_zS_z + 2h_x(t) S_x$
(here $S_{x,y,z}$ are the spin component operators related to the corresponding
Pauli matrices, e.g. \cite{ll3}). 
For the moment we assume that $|h_x(t)| \ll 2$ 
and the frequency
$\omega = 2\pi / T \ll 2$. In that case we can use the adiabatic
approximation and neglect Zener transitions. The two eigenvalues of $H$
for a given value of $h_x$ are $\lambda_{\pm} = \pm \sqrt{1+h_x^2}$.
The expectation value
for $S_x$ in these states is given by 

\begin{equation}
\langle S_x \rangle  = \frac{h_x}{2\sqrt{1+h_x^2}}\;\;.
\label{1}
\end{equation}
Now we assume that the spin is in any of the two states. Slow variation
of $h_x$ in time will keep the system in that state. Let us average
$\langle S_x \rangle$ over one period of oscillation. Because  
$\langle S_x \rangle$ is odd in $h_x$, we will obtain nonzero
time averages for the $x$-component of the spin if e.g. 
$\int_0^T h_x^3 {\rm d} t \neq 0$. This is possible if $h_x(t)$ 
contains several harmonics (SH),
e.g. $h_x(t)=h_1 \cos (\omega t) + h_2 \cos (2\omega t + 
\xi)$ (see also \cite{sfoyyz00}).
In that case in lowest order in $h_1,h_2$ we obtain
$\langle S_x \rangle = -\frac{3}{16}h_1^2 h_2 \cos \xi$. 
We conclude this example with stating
that it is possible to generate a nonzero average $S_x$ spin component
by applying a permanent field in $z$-direction and a
time-periodic field with SH and zero average in $x$-direction.

Let us relate the results from the example given above to
symmetry considerations. The Hamiltonian $H$ should be a periodic
function of time $H(t)=H(t+T)$. Instead of solving the time-dependent
Schr\"odinger equation, which would bring us to the analysis of
unitary Floquet matrices \cite{sc89}, 
we follow the density matrix approach, which
is suitable since we want to average over different initial conditions
and are thus facing the dynamics of mixed states. The density matrix $\rho$
satisfies the quantum Liouville equation \cite{ll3}
\begin{equation}
\frac{\partial \rho}{\partial t} = {\rm i} \left[ H,\rho \right]
-\nu (\rho - \rho _{\beta} )
\label{2-1}
\end{equation}
where $\left[ A,B \right] = AB-BA$, $\rho _{\beta}$ is some equilibrium
density matrix parametrized by the inverse temperature $\beta$ and
$\nu$ is a phenomenological parameter measuring the coupling
strength of the system described by $H$ to some environment. Note that
$\nu$ is the characteristic inverse relaxation time of $H$
in the environmental bath. 

Let us further define $H_0 = 1/T \int_0^T H(t) {\rm d}t$ and $H_1(t)
\equiv H(t) - H_0$. Note that $\int_0^T H_1(t) {\rm d}t =0$. 
Then we may choose $\rho_{\beta} = \frac{1}{Z} {\rm e}
^{-\beta H_0}$ with $Z = {\rm Tr} ({\rm e} ^{-\beta H_0})$. 
We define the value 
$\bar{A}(t)$
of an observable characterized
by the operator $A$ as $\bar{A}(t) = {\rm Tr} (A \rho(t))$. The time average
of $\bar{A}(t)$ shall be defined as $\tilde{A} = 
\lim_{t' \rightarrow \infty} \frac{1}{t'}\int_0^{t'} \bar{A}(t) {\rm d}t$.
The averaged attenuation power (the rate of energy transfer from the
time-periodic field to the heat bath) is given by $W = \nu (\tilde H_0 - 
{\rm Tr}(H_0 \rho_{\beta}))$.

We chose the relaxation term in (\ref{2-1}) in an oversimplified
form. There are many theories which exploit different concrete relaxation
mechanisms (e.g. \cite{Garanin} and references therein). The reason
for choosing  (\ref{2-1}) instead is that
it
allows to discuss the following symmetry breaking without entering the
details of the concrete dissipation mechanism. In other words, we deliberately 
choose the simplest dissipation term which conserves all symmetries
of our dynamical system except time reversal.

Equation (\ref{2-1}) is a linear equation for the matrix coefficients
of $\rho$ with inhomogeneous terms due to $\rho_{\beta}$. The general
solution is given by a sum of the general solution of the homogeneous equation
(put $\rho_{\beta}=0$ in (\ref{2-1})) and a particular solution of the
full equation. Since the homogeneous solution for $\nu=0$ is given by
some unitary time evolution, $\nu > 0$ will cause all solutions of
the homogeneous equation to decay to zero for infinite time. 
For $t \gg 1/\nu$ any
particular solution of the 
inhomogeneous equation
trends to a unique time-periodic solution - the attractor of
(\ref{2-1}).
This allows us to choose any (reasonable) initial condition $\rho (t=0)$.
If $H$, $\rho (t=0)$ and
$\rho_{\beta}$ are invariant under certain unitary transformations,
it immediately follows that $\rho (t)$ keeps those symmetries, and
consequently
the attractor will have the same symmetries too.
For large temperatures $\rho_{\beta}$ is approaching
the unity matrix (up to some factor). 
Consequently in that limit, whatever
the time dependence of $H(t)$, the solution of (\ref{2-1}) will approach
$\rho_{\beta}$. Finally we note that due to ${\rm Tr} \rho_{\beta}=1$
any choice of $\rho(t=0)$ with ${\rm Tr} \rho (t=0) = 1$ implies ${\rm Tr} \rho (t) = 1$ for
all $t$. 

Let us consider (\ref{2-1}) for 
\begin{equation}
H=h_0 S_z + h(t) (\alpha S_x + \gamma S_z)
\label{2-2n}
\end{equation}
where $\alpha= \sin (\phi)$ and $\gamma = \cos (\phi)$. 
This model describes a spin in a constant magnetic field pointing
in the $z$-direction, under the influence of an additional time-periodic
field $h(t)=h(t+T)$. This oscillating field should have zero mean:
$\int_0^T h(t) {\rm d}t=0$. 
Let us define $h(t)$ having $T_a$ symmetry if $h(t) = -h(-t)\equiv h_a(t)$,
$T_s$ symmetry if $h(t)=h(-t)\equiv h_s(t)$, and $T_{sh}$ symmetry 
if $h(t)=-h(t+T/2)\equiv h_{sh}(t)$
(note that in the two first cases any argument shift is allowed, so that
e.g. $h(t)=\cos (t+\mu)$ posesses all three symmetries).
For a monochromatic field (MCF) $h(t)$ and
$\phi = \pi /2$ (\ref{2-2n}) is the classical setup for performing
magnetic resonance (MR) experiments \cite{cps90},\cite{more}. 

For the 
$s=\frac{1}{2}$ case the spin component operators are given by the
Pauli matrices: $S_{x,y,z} = \frac{1}{2}\sigma_{x,y,z}$. The density
matrix $\rho$ has three independent real variables. Using the variables
$\bar{S}_{x,y,z}$ we find
\begin{eqnarray}
 \dot{\bar{S}}_x = (h_0 + \gamma h(t))\bar{S}_y -\nu \bar S _x 
 \label{2-3n} \\
 \dot{\bar{S}}_y = \alpha h(t) \bar S_z - (h_0 + \gamma h(t))
 \bar S_x - \nu \bar S_y
 \label{2-4n} \\
 \dot{\bar{S}}_z = -\alpha h(t) \bar S_y - \nu (\bar S_z - C)
 \label{2-5n}
 \end{eqnarray}
 where $C=1/2 \tanh (h_0 \beta /2)$. Note that the obtained set of equations 
 for $\nu=0$ 
 is equivalent to the Heisenberg equations for the operators $S_{x,y,z}$
 and thus also to the equations of motion for a classical spin.
 In fact (\ref{2-3n})-(\ref{2-5n}) is a particular case of the 
 Bloch equations \cite{cps90},\cite{Bloch}.
 
Let us discuss the symmetries of (\ref{2-3n})-(\ref{2-5n}) which
conserve $H_0$, i.e. $\bar S_z \rightarrow S_z$.  
Consider the case $\gamma =0$: 
if $h(t)\equiv h_{sh}(t)$ then a
symmetry operation $Q_1$ is 
$
\bar S _x \rightarrow - \bar S _x\;,\;
\bar S _y \rightarrow - \bar S _y\;,\; \bar S _z \rightarrow  \bar S _z
\;,\;
t \rightarrow t + T/2\;\;$.
If $Q_1$ holds we conclude that $\tilde S _x = \tilde S _y =0$,
while $\tilde S_z$ may be nonzero.
Consider $\gamma=0$ and $\nu=0$:
if $h(t) \equiv h_a(t)$ then a symmetry
operation $Q_2$ is  
$
\bar S _x \rightarrow - \bar S _x\;,\;
\bar S _y \rightarrow  \bar S _y\;,\; \bar S _z \rightarrow  \bar S _z
\;,\; t \rightarrow  - t \;\;$.
If $Q_2$ holds it follows $\tilde S _x =0$, while $\tilde S_{y,z}$ 
may be nonzero.
Finally for $\nu=0$ and $h(t) \equiv h_s(t)$ a symmetry operation
$Q_3$ is
$
\bar S _x \rightarrow  \bar S _x\;,\;
\bar S _y \rightarrow  - \bar S _y\;,\; \bar S _z \rightarrow  \bar S _z
\;,\; t \rightarrow  - t \;\;$. If $Q_3$ holds it follows
$\tilde S_y=0$, while $\tilde S_{x,z}$ may be nonzero.

Let us note some consequences. If we choose $h(t)=h_1 \cos (\omega t)$,
then the classical MR setup with $\gamma =0$ ($Q_1$) yields nonzero values
for $\tilde S_z$ only \cite{cps90}.
If the probing field is not perpendicular to the $z$-axis ($\gamma \neq 0$),
nonzero values for $\tilde S_x$ and $\tilde S_y$ appear as well.
$\tilde S_y$ will vanish in the limit of zero coupling to the
environment $\nu \rightarrow 0$ ($Q_3$), so
that this average can be used to measure the coupling strength.
Applying e.g. $h(t)=h_1 \sin (\omega t) + h_2 \sin( 2\omega t)$
(having $h_a$ symmetry but not $h_{sh}$ and $h_s$ one) we can suppress the
value of $\tilde S_x$ relatively to $\tilde S_y$ for $\gamma \rightarrow 0$ and
$\nu \rightarrow 0$
keeping $\tilde S_y$ finite ($Q_2$)!

Analytical solutions to (\ref{2-3n})-(\ref{2-5n}) can be found e.g.
for large $\nu \gg 1$. Expanding in $1/\nu$ and averaging over time
we find in lowest orders
\begin{eqnarray}
\tilde S_x = C \alpha \gamma \langle h^2\rangle  \frac{1}{\nu ^2} -
C \alpha  (-\gamma \langle h \ddot{h} \rangle  +3\gamma h_0^2 \langle h^2\rangle + 
\nonumber \\
(\alpha^2 + 3\gamma^2) h_0 \langle h^3\rangle +
\gamma (\gamma^2 + \alpha ^2)\langle h^4\rangle ) \frac{1}{\nu ^4} + O(\frac{1}{\nu ^5}) 
\label{2-6n} \\
\tilde S_y = -C \alpha \left[ 2\gamma h_0 \langle h^2\rangle  + (\gamma ^2
+ \alpha^2) \langle h^3\rangle  \right] 
\frac{1}{\nu ^3} + O(\frac{1}{\nu ^5})
\label{2-7n}
\end{eqnarray}
where $\langle f(t)\rangle  = \frac{1}{T}\int_0^T f(t) {\rm d}t$. It is easy to
cross check that all symmetry statements from above are correct.
Nonzero values for $\langle h^3\rangle $ can be obtained e.g.
with $h(t)=h_1 \sin (\omega t) + h_2 \sin( 2\omega t+ \xi)$ for
$\xi \neq 0,\pi$ (see also \cite{sfoyyz00}). 

In Fig.1 we show the dependence of $\tilde S_{x,y,z}$ on $\omega$
for $h(t)=\sqrt{2} \cos \omega t$, $\phi=\pi/4$, $h_0=3$, $\nu=0.1$
and $\beta =10$.
The time-periodic field has a large amplitude compared to typical
MR setups \cite{cps90}. This causes the $\tilde S_z$ curve to show
a rather broad peak at $\omega \approx h_0$ - the position of
the expected MR resonance. However we also observe sattelite peaks
at lower frequencies which are clearly related to the variations
of nonzero $\tilde S_{x,y}$ (for convenience these averages
are scaled by a factor of 10 in Fig.1). In fact the positions
of the sattelite peaks are subharmonics of the main resonance. 
The dependence of $\tilde S_x$
and $\tilde S_y$ on $\omega$ shows rather complex structures. We find that
typically the dependence of these averages on $\omega$ becomes oscillatory
for small $\omega \ll h_0$, whereas large $\omega$ values yield
smooth decay curves. Note also that these averages stay nonzero down
to small frequencies in accord with the adiabatic example
from above. Also important is to notice that the fluctuations
of $\bar S_x$ and $\bar S_y$ around their mean values may happen with
amplitudes being one order of magnitude larger than the mean values
(see inset in Fig.1 ).

The above results hold also for larger spins. To show that they
also hold for internal anisotropy fields rather than external
fields, we consider a spin with $s=1$ and the Hamiltonian 
\begin {equation}
H = S_z^2 + h(t) (\alpha S_x +\gamma S_z) \label{2-2}
\end{equation}
which describes a spin with an anisotropy along the $z$-axis ($S_z^2$)
under the influence of an external 
magnetic field $h(t)$ parallel to the $xz$ plane. The magnetic field
is again time-periodic with period $T$ and has zero mean.
The $3\times 3$ hermitian density matrix $\rho$ has 8 independent
real parameters. 
Since $H$ in (\ref{2-2}) is a real symmetric matrix, we can 
define $\rho= R + {\rm i}I$ where $R$ is a real symmetric matrix
and $I$ a real antisymmetric one. Noting that also $\rho_{\beta}$ is
a real diagonal matrix, (\ref{2-1}) can be rewritten as
\begin{eqnarray}
\frac{\partial R}{\partial t} = - \left[ H,I \right]
-\nu (R - \rho _{\beta} ) \label{2-8n} \\
\frac{\partial I}{\partial t} =  \left[ H,R \right]
-\nu I
\label{2-9n}
\end{eqnarray}
It follows $\bar S_x= \sqrt{2}(R(1,2)+R(2,3))$, 
$\bar S_y = -\sqrt{2}(I(1,2)+I(2,3))$ and
$\bar S_z = R(1,1) - R(3,3)$ \cite{Matrix}.
Using the abbrevations $P_x=\sqrt{2}(R(1,2)-R(2,3))$, 
$P_y=\sqrt{2}(I(1,2)-I(2,3))$, $P_z=R(1,1)+R(3,3)$,
$R_{13}=\sqrt{2}R(1,3)$, $I_{13}=\sqrt{2}I(1,3)$, 
$R_{22}=\sqrt{2}R(2,2)$, $D^{-1}=1+2{\rm e}^{-\beta}$ and
$F^{-1}=2+{\rm e}^{\beta}$  the equations of motion become
\begin{eqnarray}
\dot{\bar{S}}_x = -P_y + \gamma h \bar S_y - \nu \bar S_x
\nonumber \\
\dot{P}_x = \bar S_y - \gamma h P_y + \sqrt{2} \alpha h I_{13} - \nu P_x
\nonumber \\
\dot{\bar{S}}_y = - P_x - \gamma h \bar S_x + \alpha h \bar S_z - 
\nu \bar S_y
\nonumber \\
\dot{P}_y = \bar S_x + \gamma h P_x + \alpha h \left[ \sqrt{2} R_{22}
-P_z - \sqrt{2} R_{13} \right] - \nu P_y
\nonumber \\
\dot{\bar{S}}_z = \alpha h \bar S_y - \nu \bar S_z  
\label{2-10n} \\
\dot{P}_z = \alpha h P_y - \nu \left[ P_z - 
2F \right]
\nonumber \\
\dot{R}_{13} = -2\gamma h I_{13} + \sqrt{2}\alpha h P_y - \nu R_{13}
\nonumber \\
\dot{I}_{13} = 2 \gamma h R_{13} - \sqrt{2} \alpha h P_x - \nu I_{13}
\nonumber \\
\dot{R}_{22} = \sqrt{2} \alpha h P_y - \nu \left[ R_{22} - \sqrt{2}
D \right]
\nonumber 
\end{eqnarray}
These equations conserve the trace
${\rm Tr} \rho \equiv P_z + R_{22}/\sqrt{2} = 1$.

Now we can discuss the symmetries of (\ref{2-10n}) which change the sign
of $\bar S$. Two of them hold only for
$\nu=0$. First, if $h(t)\equiv h_a(t)$,
then the equations are invariant under change of sign of the variables
$t,\bar S_x, \bar S_y, \bar S_z$ (leaving all other variables unchanged).
A second case takes place if $h(t) \equiv h_s(t)$. Then changing the sign
of $t,\bar S_y, P_y, I_{13}$ (leaving all other variables unchanged)
is an operation which keeps equations (\ref{2-10n}) invariant.
These two cases imply that if $h(t)$ is antisymmetric, then for 
vanishing dissipation $\nu \rightarrow 0$ $\tilde S_{x,y,z} \rightarrow 0$,
while for symmetric $h(t)$ the same limit provides a vanishing of
the $y$-component only $\tilde S_y \rightarrow 0$.

For the general case $\nu \neq 0$ two more symmetries may take place.
If $\gamma=0$ (the field $h(t)$ acts perpendicularly to the anisotropy
axis $z$), changing the sign of $\bar S_y, \bar S_z, P_x, I_{13}$
(and keeping all others) leaves (\ref{2-10n}) invariant. Finally if
$h(t)\equiv h_{sh}(t)$, the shift $t \rightarrow t+T/2$ and simultaneous
change of sign of the variables $\bar S_x, \bar S_z, P_y, I_{13}$
do not change the equations. It follows that $\tilde S_y=
\tilde S_z = 0$ for $\gamma=0$ and $\tilde S_x=\tilde S_z=0$ for
$h(t)$ having shift symmetry. 

It is interesting to note that for a MCF 
$h(t)=\cos \omega t$ and $\nu \neq 0$, $\gamma \neq 0$ the spin will
point on average in $y$ direction, i.e. perpendicular to the plane
spanned by the driving field and the local anisotropy axis!
In Fig.2 we plot the dependence of $\tilde S_{y}$ on $\omega$
for this case ($\beta = 10$, $\nu = 0.1$, $\gamma = \alpha =1$),
which confirms the symmetry considerations. Note that $\tilde S_x$ and
$\tilde S_z$ are less than $10^{-8}$ as found in the numerical studies.

To conclude this case we remark that it is again an easy task to perform
expansions in $1/\nu$ for large $\nu$ values as shown above for the
$s=1/2$ case. The resulting expressions also confirm the symmetry
considerations.

So far we have discussed the results for quantum spin systems.
It is also possible to analyze corresponding classical
systems. E.g. the classical equations for (\ref{2-2})
are given by
\begin{eqnarray}
\dot{s}_x = -2s_zs_y + \gamma h s_y \;\;\label{3-2} \\
\dot{s}_y = 2s_zs_x + h\left( \alpha s_z - \gamma s_x \right)
\label{3-3} \\
\dot{s}_z = -\alpha h s_y \label{3-4}
\end{eqnarray}
Let us discuss the symmetry properties of (\ref{3-2})-(\ref{3-4}).
We denote on the left part the condition and on the right part the
symmetry operations which leave the equations of motion invariant
(note that we list only those variables which have to be changed):
\begin{eqnarray}
\gamma=0  & :\;\;\; & (s_y\;,\;s_z) \rightarrow (-s_y\;,\;-s_z)
\label{3-5} \\
T_a & :\;\;\; & t\rightarrow -t\;,\;(s_x\;,\;s_y\;,\;s_z) 
\rightarrow (-s_x\;,\;-s_y\;,\;-s_z)
\label{3-6} \\
T_s & :\;\;\; & t \rightarrow -t \;,\; s_y \rightarrow -s_y 
\label{3-7} \\
T_{sh} & :\;\;\; & t \rightarrow t+\frac{T}{2}\;,\;
(s_x\;,\;s_z) \rightarrow (-s_x\;,\;-s_z)
\label{3-8}
\end{eqnarray}
If we add dissipation terms, these terms will break time reversal
symmetry, and we are left only with (\ref{3-5}) and (\ref{3-8}).
All of the above statements for the quantum system can be recovered.
Especially nonzero dissipation and $\gamma,\alpha \neq 0$ lead to nonvanishing
magnetization along the $y$-axis, even for MCF.

Let us summarize the presented results. We have shown that
time-periodic magnetic fields with zero mean may induce nonzero
averages of spin components which would be strictly zero in the absence
of these fields. The spin is simultaneously experiencing some 
local anistropy field or simply an external constant field. 
In addition the spin is coupled to some thermal environment
characterized by some finite temperature and a characteristic
relaxation time \cite{Retardation}.
The
reasoning follows symmetry considerations of the dynamical equations.
In the case of a classical spin these equations formally coincide
with the Heisenberg equations for the quantum spin operators. In the quantum
case we instead solve the (purely linear!) equations of motion for
the independent components of the density matrix. Remarkably the symmetry
properties obtained from both approaches coincide.  

The quantum approach shows that for infinite temperatures all spin component
averages will vanish. 
This follows from $\rho(t \rightarrow \infty;\beta \rightarrow 0) 
= \rho_{\beta \rightarrow 0}$
and ${\rm Tr} S_{x,y,z} = 0$.

For the spin $1/2$ case we proposed a MR experiment to observe
the effect. One should choose the time-periodic magnetic field to be not
perpendicular to the static magnetic field. Further the amplitude
of the time-periodic field should be not too small such that the
generated $\tilde S_x$ and $\tilde S_y$ components are measurable.
The attenuation spectrum should show resonances located at subharmonics
of the original resonance. 
The intensity of the sattelite peaks is a function of both
the angle between both fields and the inverse relaxation time $\nu$.

Experiments which probe directly the nonzero spin components 
can be performed by adding yet another probing field to the system,
and varying its frequency while keeping the frequency of the
original probing field. This will be studied in detail in future work.
\\
\\
Acknowledgements.
\\
We thank A. Bartl, P. Fulde, D. A. Garanin and A. Latz for useful discussions.

\newpage

\noindent
Figure captions.
\\
\\
Fig.1. $10\tilde S_{x}$ (solid), $10\tilde S_y$ (dashed) and
$\tilde S_z$ (dotted) as  functions of $\omega$ (see text for
parameters). 
\\
Inset: $\bar S_{x,y,z}$ versus time for one period of $h(t)$
at $\omega=1.5$ (same line codes as in Fig.1). Note that functions
are not scaled here!
\\
\\
Fig.2.
\\
\\
$\tilde S_y$ as a function of $\omega$ (see text for
parameters).


\begin{thebibliography}{10}
\bibitem{Bloch}
Our equations (\ref{2-3n})-(\ref{2-5n}) differ
from the corresponding Bloch equations since we did not choose different
relaxation times for the different spin components. 
However choosing different relaxation times would not
change the symmetry properties discussed in the paper, and consequently
a variation of these relaxation times would cause only a quantitative
change.
\end{thebibliography}
\end{document}